\documentstyle[aps]{revtex}

\begin{document}
\makeatletter
\title{Space of the vertices of relativistic spin networks}
\author{A. Barbieri}
\address{Dipartimento di Fisica dell'Universit\`a di Pisa\\ 
P.zza Torricelli 2, I-56100 Pisa, Italy\\
INFN-Sezione di Pisa\\
{\rm E-mail:abarbier@ibmth.difi.unipi.it}}
\makeatother
\maketitle
\begin{abstract}
\\The general solution to the constraints that define relativistic spin 
neworks vertices is given and their relations with 3-dimensional quantum 
tetrahedra are discussed. An alternative way to handle the constraints is 
also presented.
\end{abstract}
\section{introduction}
The purpose of this note is to 
review some aspects of the extension to four dimensions 
\cite{barcr,baez} of the quantization procedure of 3-simplices
introduced in \cite{io} for the 3-dimensional case.
In \cite{barcr,baez} this extension is used to build a quantum 4-simplex, 
but since the algebraic structure we are going to explore can be dealt with 
also in the ``reduced'' context of 3-simplices, we shall not be concerned 
with this further step.
In section \ref{quant} we show that 
the general solution to the constraints that define the quantum 3-simplex 
in four dimensions is given by the single solution found by Barrett and 
Crane.\\
This result is somewhat disappointing since the space of states of 
a quantum tetrahedron, being described in terms of {\em intrinsic} geometry, 
should be the same independently from the dimensionality of the space in which 
it is embedded. In section \ref{weak} we discuss an alternative way to 
handle the constraints, following which one obtains a space isomorphic to 
that of the 3-dimensional case.

We warn the reader that the results announced depend crucially on two
conjectures:
\begin{description}
\item[{\rm a)}] the operator ${\bf U}$ has a non degenerate spectrum in 
${\cal H}^0_{\{J\}}$ (see \cite{io} for the notations used in the paper);
\item[{\rm b)}] nontrivial $6j$-symbols are never zero.
\end{description}
\section{Quantum tetrahedra}\label{quant}
The basic idea is to use as fundamental variables the 
bivectors defined by 2-dimensional faces of the tetrahedron rather than the 
edges. Since the set of bivectors in {\it n} dimensions is isomorphic to the 
dual of the Lie algebra of $SO(n)$, we may define the Hilbert space of a 
quantum bivector as the orthogonal sum of the inequivalent irreducible 
representations of the universal covering of $SO(n)$ (see \cite{baez} 
for a deeper discussion); if $n=3$ the covering group is $SU(2)$, 
while if $n=4$ it is {\it Spin}$(4)\simeq SU(2)\otimes SU(2)$. 
In the following we will focus on these two subcases.

The next step consists in taking the tensor product of four such spaces (one 
for each face) and imposing some constraints which guarantee that the faces 
have suitable properties, which, translated in terms of bivectors, read: 
\begin{enumerate}
\item the bivectors defined by the faces must (when oriented properly) add 
up to zero;
\item the bivectors are all simple (i.e.\@ they can be written as wedge 
products of two vectors);
\item the sum of two bivectors is also simple.
\end{enumerate}
In three dimensions all bivectors are simple and the only constraint 
is that the faces close: the resulting space is that of 4-valent 
$SU(2)$-spin network vertices.\\ 
In four dimensions there are non-simple bivectors and we must handle all the 
constaints.

If we denote the bivectors by $n(f)$, where the index $f$ labels the 
faces, the constraints assume the following form:
\begin{eqnarray}
\sum_f\sigma(f)n(f) &= & 0,\label{uno}\\
\langle n(f),*n(f)\rangle &= & 0,\label{due}\\
\langle n(f),*n(f')\rangle &= & 0,\label{tre}
\end{eqnarray}
where $\sigma(f)=\pm 1$ takes into account the orientation properties 
and $*$ denotes the duality operator on bivectors.

When we quantize bivectors, i.e.\@ we label each face with a pair of spins,
imposing eqn. (\ref{uno}) is equivalent to confine oneself in the space
\begin{equation}
^{(4)}\overline{\cal H}{}_\tau\equiv\bigoplus_{\{\rho\}}
{\rm Inv}[\bigotimes_{f\in\partial\tau}\rho(f)],
\end{equation}
where $\{\rho\}$ ranges over all labelings of the 2-dimensional faces in 
$\tau$ (an abstract 3-simplex) with pairs of spins: this is the 
space of 4-valent {\it Spin}(4)-spin networks vertices.

Condition (\ref{due}) is equivalent to the imposition that the 
self- and antiself-dual parts of the bivector are of equal norm and 
can be translated in the quantum context to the restriction to 
representations $\rho$ of the form $(j,j)$.

Let us now turn to the last constraint (\ref{tre}): the quantum analogue is 
\begin{equation}
[{\bf n}^{+}_{f\!f'}-{\bf n}^{-}_{f\!f'}]
|\psi\rangle\equiv{\bf O}_{f\!f'}|\psi\rangle=0\label{sempl}
\end{equation}
on the space $^{(4)}{\cal H}_\tau$ of the quantum tetrahedron (the signs $\pm$ 
denote the self- and antiself-dual $SU(2)$-sectors of {\it Spin}(4)). 
An equation of this kind must hold
for every couple of faces $f$ and $f'$ so that the operators 
${\bf O}_{f\!f'}$ and ${\bf O}_{f'\!f''}$ must commute on 
$^{(4)}{\cal H}_\tau$. 
Since
\begin{equation}
[{\bf O}_{f\!f'},{\bf O}_{f'\!f''}]=\pm{\rm i}[{\bf U}^+ +{\bf U}^-],
\end{equation}
where the overall sign depends on the ordering chosen to define ${\bf U}$,
it is clear that $^{(4)}{\cal H}_\tau$ must be contained in the space
$${\rm DIAG}_{{\bf U}}\equiv\bigoplus_{\{J\},m}|\{J\},m\rangle_+
\otimes|\{J\},-m\rangle_-,$$
where $\{J\}$ ranges over all labelings of the faces with {\it a single} spin
in such a way that one can obtain a 4-valent $SU(2)$-spin network vertex and 
$m$ runs over the spectrum of ${\bf U}$ in ${\cal H}^0_{\{J\}}$.
This space is well defined because of conjecture a) and because 
${\bf U}$ has symmetric eigenvalues with respect to zero.

Eqn.\@ (\ref{sempl}) is equivalent to the following one (we now work with 
fixed $\{J\}$):
\begin{equation}
(\langle m'|_+\otimes\langle -m''|_-){\bf O}_{f\!f'}|\psi\rangle=0\ \ \ 
\forall(m',m'').
\label{strong}
\end{equation}
Since we are looking for $|\psi\rangle\in{\rm DIAG}_{{\bf U}}$, we 
may write
$$|\psi\rangle=\sum_m c_m|m\rangle_+
\otimes|-m\rangle_-$$
and eqn.\@ (\ref{strong}) becomes 
\begin{equation}
0=\sum_mc_m[\delta_{mm''}\langle m'|{\bf n}_{f\!f'}|m\rangle
-\delta_{mm'}\langle -m''|{\bf n}_{f\!f'}|-m\rangle]=c_{m''}
\langle m'|{\bf n}_{f\!f'}|m''\rangle-c_{m'}
\langle -m''|{\bf n}_{f\!f'}|-m'\rangle.
\label{delte}
\end{equation}
We may distinguish two cases: $m'=m''$ and $m'\neq m''$.\\
If $m'=m''$ eqn.\@ (\ref{delte}) gives
\begin{equation}
0=c_m[\langle m|{\bf n}_{f\!f'}|m\rangle-
\langle -m|{\bf n}_{f\!f'}|-m\rangle]\ \ \ \forall m;\label{weakc}
\end{equation}
since however ${\bf n}_{f\!f'}$ commutes with the ``parity'' operator ${\bf P}$
defined in \cite{io}, the two matrix elements in the r.h.s.\@ are equal so 
that this equation is always satisfied.\\ 
If conversely $m'\neq m''$, the equation becomes
$$0=c_{m''}\langle m'|{\bf n}_{f\!f'}|m''\rangle-c_{m'}
\langle -m''|{\bf n}_{f\!f'}|-m'\rangle.$$
It is now crucial to observe that, since the ``parity'' operation is supposed 
to reverse the normals, which in last instance are angular momenta, it 
must be an {\em antiunitary} operator (i.e.\@ it is actually the operator 
implementig time-reversal, from the ``spin'' point of view). Using 
${\bf P}|m\rangle=|-m\rangle$, $[{\bf n}_{f\!f'},{\bf P}]=0$
and the antilinearity of ${\bf P}$ it is not difficult to show that
$$\langle m'|{\bf n}_{f\!f'}|m''\rangle=
\langle -m''|{\bf n}_{f\!f'}|-m'\rangle,$$
so that eqn.\@ (\ref{strong}) becomes
$$0=(c_{m'}-c_{m''})\langle m'|{\bf n}_{f\!f'}|m''\rangle,$$
which admits the nontrivial solution 
\begin{equation}
|\psi\rangle={\rm cost}\sum_m|m\rangle_+\otimes|-m\rangle_-.
\label{soluz}
\end{equation} 
The only escape to the conclusion that this is the only one, is that there 
are subspaces spanned by some of the $|m\rangle$ that are invariant under 
the action of ${\bf n}_{f\!f'}$. However, since eqn.\@ 
(\ref{strong}) must hold for every couple of faces, such subspaces 
must be invariant under the action of {\em all} the operators 
${\bf n}_{f\!f'}$.
The only candidate subspaces with this property are those spanned by 
eigenvectors of ${\bf n}_{f\!f'}$ and, since we have e.g.\@
$$\langle j'{}\!\!_{f\!f'}|{\bf n}_{f'\!f''}|j_{f\!f'}\rangle\neq 0\Rightarrow
j'{}\!\!_{f\!f'}-j_{f\!f'}=0,\pm 1,$$
if it were one of them it would be possible to find orthogonal eigenvectors 
of ${\bf n}_{f\!f'}$ and ${\bf n}_{f'\!f''}$; but the identity
$$\langle j_{f\!f'}|j_{f'\!f''}\rangle=(-1)^{\sum_f j(f)}
\sqrt{(2j_{f\!f'}+1)(2j_{f'\!f''}+1)}\left\{\matrix{j(f) & j(f') & j_{f\!f'} 
\cr j(f'') & j(f''') & j_{f'\!f''}}\right\},$$
shows that such an eventuality violates conjecture b), so that we are left 
with the only solution  
(\ref{soluz}), which should be the one found by Barrett and Crane.
\section{Weak constraint}\label{weak}
It seems that, for fixed external spins, the space of states of a quantum 
tetrahedron depends on the dimensionality of the space in which it is 
embedded: a strange result, being the whole construction made in terms
of {\em intrinsic} geometry. A result that sounds even odder if we consider 
that there are values of the external spins for which the mean geometries 
computed in \cite{io} {\em depend} on the eigenvalues of ${\bf U}$, so that 
the mean geometry computed in the state (\ref{soluz}) would be different 
and not within a scale factor (see below).

The fact that eqn.\@ (\ref{weakc}) is identically satisfied suggests however
a very simple way to overcome this ``strangeness'' which consists in 
imposing the last constraint only in a ``weak'' sense, i.e.\@ only to the 
matrix elements between ``physical states''. In this sense the space 
of solutions is the entire ${\rm DIAG}_{{\bf U}}$ which indeed has the same 
dimensions of the Hilbert space of a quantum tetrahedron in three dimensions.
It is worth noting that the matrices representing all the nonchiral operators 
$${\bf N}_{f\!f'}\equiv {\bf n}^+_{f\!f'}+{\bf n}^-_{f\!f'},$$
are {\em diagonal} in this space, so that the states 
$|m\rangle_+\oplus|-m\rangle_-$ may be regarded as {\em eigenstates of 
geometry} and the ``igengeometries'' are proportional to the mean geometries 
computed in \cite{io}.\\  
Indeed, with the same normalization conventions (symmetry 
factor present in the formula for the norm of a bivector but not in the 
wedge products), the eigenvalues of the ``squared area'' operators 
get a multiplicative factor 4 with respect to the 3-dimensional case 
and the edge lenghts are all multiplied by a factor 
$\sqrt{2}$. It is interesting to observe that, regardless the kind
of constraint imposed, the spectra of these operators come to agree with 
those computed in the Loop Representation framework of Quantum Gravity 
(ignoring the Immirzi parameter) in the units of \cite{deprov}.

We leave to future investigations the search for the meaning of these two 
ways to handle the constraints.
\acknowledgments
I thank John Baez for his invaluable observations about the previous 
versions of this paper.  

\thebibliography{99}
\bibitem{barcr} J.W.\@ Barrett, L.\@ Crane, {\it Relativistic spin networks
and quantum gravity}, preprint available as {\bf gr-qc/9709028} (1997)
\bibitem{baez} J.C.\@ Baez, {\it Spin Foam Models}, preprint available as 
{\bf gr-qc/9709052} (1997)
\bibitem{io} A.\@ Barbieri, {\it Quantum tetrahedra and simplicial spin 
networks}, preprint available as {\bf gr-qc/9707010} (1997)
\bibitem{deprov} R.\@ De Pietri, C.\@ Rovelli, Phys.\@ Rev.\@ {\bf D54},
2264 (1996)
\end{document}